\documentclass[aps,prl,amsmath,amssymb,superscriptaddress,showkeys,reprint]{revtex4-1}
\usepackage{graphicx}
\usepackage{dcolumn}
\usepackage{bm}
\usepackage{CJKutf8}
\usepackage{natbib}
\usepackage[utf8]{inputenc}
\usepackage{hyperref}
\begin{document}

\title{Optomechanical Transduction and Characterization of a Silica Microsphere Pendulum via Evanescent Light}

\author{Ramgopal Madugani}
\affiliation{Light-Matter Interactions Unit, Okinawa Institute of Science and Technology Graduate University, Onna, Okinawa 904-0495, Japan}
\affiliation{Physics Department, University College Cork, Cork, Ireland}

\author{Yong Yang(\begin{CJK*}{UTF8}{gbsn}杨 勇\end{CJK*})}
\affiliation{Light-Matter Interactions Unit, Okinawa Institute of Science and Technology Graduate University, Onna, Okinawa 904-0495, Japan}
\affiliation{National Engineering Laboratory for Fiber Optics Sensing Technology, Wuhan University of Technology, Wuhan, 430070, China}

\author{Jonathan M Ward}
\affiliation{Light-Matter Interactions Unit, Okinawa Institute of Science and Technology Graduate University, Onna, Okinawa 904-0495, Japan}

\author{Vu H Le}
\affiliation{Light-Matter Interactions Unit, Okinawa Institute of Science and Technology Graduate University, Onna, Okinawa 904-0495, Japan}

\author{S\'{\i}le Nic Chormaic}
\email{sile.nicchormaic@oist.jp}
\affiliation{Light-Matter Interactions Unit, Okinawa Institute of Science and Technology Graduate University, Onna, Okinawa 904-0495, Japan}

\date{\today}

\begin{abstract}
Dissipative optomechanics has some advantages in cooling compared to the conventional dispersion dominated systems. Here, we study the optical response of a cantilever-like, silica, microsphere pendulum, evanescently coupled to a fiber taper. In a whispering gallery mode resonator the cavity mode and motion of the pendulum result in both  dispersive and dissipative optomechanical interactions. This unique mechanism leads to an experimentally observable, asymmetric response function of the transduction spectrum which can be explained using coupled-mode theory. The optomechanical transduction, and its relationship to the external coupling gap, are investigated and we show that the experimental behavior is in good agreement with the theoretical predictions. A deep understanding of this mechanism is necessary to explore trapping and cooling in dissipative optomechanical systems. 
\end{abstract}


\maketitle

Cavity optomechanics has witnessed rapid progress in recent years and has been implemented in many systems, such as Fabry-P\'erot (F-P) microcavities, photonic crystal (PhC) devices, and whispering gallery mode (WGM) microcavities \cite{Kippenberg:07,Aspelmeyer2013}. Cavity optomechanics is based on the interaction between the cavity's photon state and the phonons of the cavity's mechanical mode; for this reason cavity optomechanical systems exhibit great potential in many applications, especially in quantum physics. Various related experiments have been done, such as resolved sideband cooling\cite{Schliesser2009,Park2009}, strong coupling\cite{Groblacher2009,Verhagen2012} and optomechanical induced transparency\cite{Weis2010,Dong2012}. Among applications, cooling the mechanical mode to its motional ground state is a crucial step for further revealing its quantum nature.

Generally speaking, if we consider the F-P cavity as an example,  mechanical oscillation of the cavity's mirror shifts the optical mode resonances by changing the cavity length; this is a dispersive interaction. Almost all  optomechanical systems fall into this category. Besides dispersion, another mechanism exists, whereby the mechanical motion modulates the cavity dissipation instead of the resonance frequencies. Compared to the case of cooling using dispersive optomechanics, the resolved sideband regime is not required, so that, theoretically, even for a bad cavity, ground state cooling is possible\cite{PhysRevLett.107.213604}. In practice, nanomechanical resonators in F-P cavities have been used to show that  pure dissipative-driven optomechanics\cite{Favero2008,Favero2009} is feasible. Recently, interference between the dissipative and dispersive coupling mechanisms has been shown in an F-P structure composed of two suspended PhC reflectors\cite{Wu2014}. Such interference was used to enhance the sensitivity of an optomechanical toque sensor.

Besides optomechanics in F-P cavities, WGM resonators provide an alternative system for optomechanics. The WGM resonator is a traveling wave cavity relying on continuous total internal reflection in the dielectric material. These structures have even higher optical quality (Q) factors than F-P cavities, while back action can be achieved via the radiation pressure or optical gradient forces, or nonlinearities. Apart from  resolved sideband cooling\cite{Schliesser2009,Park2009,Verhagen2012}, other applications including feedback cooling\cite{Gavartin2012hybrid} and optomechanical induced transparency in microtoroids\cite{Weis2010} and microspheres\cite{Dong2012,Kim2015} have been performed. Due to the small mechanical displacement and the fiber taper coupling configuration, the reported optomechanics in WGM cavities are mostly dispersion dominated with negligible  contribution from dissipation. In this paper, we  provide experimental evidence that the dissipative coupling mechanism does exist in a special WGM cavity, i.e., the microsphere pendulum. Interference between the dispersion and dissipation channels in this system leads to an asymmetric transduction. Furthermore, the strength of the dissipation and dispersion can be adjusted by varying the coupling gap between the taper and the micropendulum.

\begin{figure}
\includegraphics[width=8.4cm]{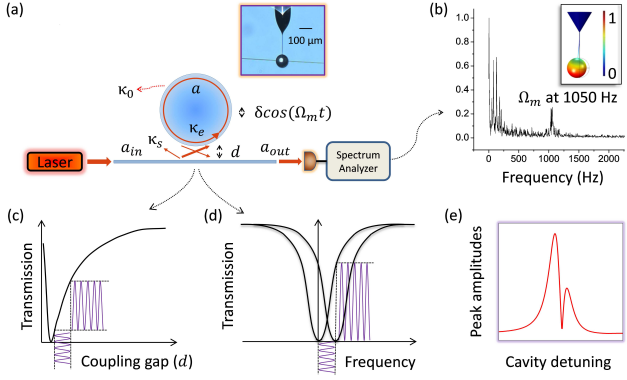}
\caption{\label{fig:ExpSetup}(a) Schematic of a taper-coupled micropendulum system. Inset: image of the micropendulum; (b) Fourier transform spectrum of the transmitted signal through the fiber taper as the pendulum moves relative to the fiber. The mechanism of such optomechanical transduction is due to the mechanical motion-induced (c) dissipative and (d) dispersive modulation. (e) A typical transduction response to cavity detuning for the 1 kHz mechanical mode peak.}
\end{figure}

A micropendulum consists of a microsphere that acts as a bob, attached to a thin stem so that it can swing freely. The microsphere supports high Q-factor whispering gallery modes. Such a simple configuration features quite low fundamental mechanical frequencies (in the Hz-kHz range) and relatively large amplitude (typically nm) oscillations. A tapered optical fiber can be used for evanescent coupling of light into the WGMs, as illustrated in Fig.\ref{fig:ExpSetup}(a). When probe light in the coupling fiber is tuned near to a WGM resonance (see Fig.\ref{fig:ExpSetup}(b)) the large amplitudes of the mechanical oscillations of the micropendulum produce a variation in the fiber transmission signal which is high above the noise level. The signal strength is related to the detuning of the laser relative to the WGM resonances. Previously \cite{Wu2012}, we showed that the signal should be stronger if the laser is red-detuned from resonance. In this paper, we will explain this effect in more detail and discuss the mechanisms involved from both theoretical and experimental viewpoints.

A coupled-mode theoretical framework can be used to describe the system depicted in Fig. \ref{fig:ExpSetup}(a). The slowly varying term of the normalized intracavity electromagnetic (EM) field, $a$, is:\cite{Spillane2003}
\begin{equation}
\frac{da}{dt}=-(\kappa_0+\kappa_s+\kappa_e+i\triangle)a+\sqrt{2\kappa_e}a_{in},
\label{eq:coupledmode}
\end{equation}
where $\triangle=(\omega_l-\omega_0)$ is the optical detuning, with $\omega_l$ and $\omega_0$ representing the laser and cavity WGM resonant frequencies, respectively. $\kappa_0$, $\kappa_s$, and $\kappa_e$ are decay rates representing the intrinsic, scattering, and external Q-factors, respectively. $\kappa_e$ and $\kappa_s$ are related to the coupling gap between the micropendulum and the tapered fiber. $\kappa_s$ is light scattered at the taper coupling region and limits the ideality of the taper-coupled system\cite{Spillane2003}. The input laser source with power, $P_{in}$, determines the input normalized amplitude, $a_{in}$, such that $a_{in}=\sqrt{P_{in}/\hbar\omega_l}$. The detected signal at the output of the fiber satisfies the input-output relationship \cite{Rasoloniaina2014} $a_{out}=-a_{in}+\sqrt{2\kappa_e}a$. The lifetime of photons in the microcavity is several tens of ns to several $\mu$s and is defined by the Q-factor, while the oscillation period of the micropendulum is usually on the ms scale. For example, a finite element analysis simulation shows that the first order mechanical mode of a micropendulum with an 88 $\mu$m sphere diameter, a 130 $\mu$m long and 2 $\mu$m  diameter stem,  is 1.05 kHz. Therefore, the steady state condition is always satisfied during the mechanical oscillation; hence, $da/dt=0$ in Eq. \ref{eq:coupledmode}. The steady state intracavity field can be expressed as $a=\sqrt{2\kappa_e}a_{in}/(i\triangle+\kappa_0+\kappa_s+\kappa_e)$. The final transmission through the fiber, $T=\left|a_{out}/a_{in}\right|^2$, is:
\begin{equation}
T=\left|1-\frac{2K_e}{(1+K_s+K_e+i\Delta)}\right|^2.
\label{eq:Trans}
\end{equation}
In Eq.\ref{eq:Trans} and all the following we associate $K_e, K_s$, and $\Delta$ as normalized parameters with $\kappa_0$. Essentially, as the pendulum oscillates, these parameters all vary with the coupling gap, $d$; hence, their values are time dependent. In an ideal optical coupling scenario, scattering effects would not appear. Hence, for the following simplified theoretical discussion, and to arrive at an explicit analytical expression comprising dispersion and dissipation, we disregard $K_s$ temporarily.

The relative motion of the fiber and the microsphere causes a frequency shift of the WGMs \cite{Gavartin2012hybrid,guo2006} and this contributes a dispersive perturbation to the detuning term such that $\Delta(t)=\Delta_0-(\Delta_g\cos(\Omega_mt))$. Here, $\Delta_0$ is the unperturbed detuning term.  Supposing that the micropendulum's displacement and frequency are $\delta$ and $\Omega_m$, respectively, then the dispersion can be defined as $\Delta_g=(\partial\omega_0/\partial d) \times\delta/\kappa_0$. The external coupling, $K_e(t)$, is defined by $K_e(t)=K_e^0\exp(\gamma\delta \cos(\Omega_mt))$, where $\gamma$ is a coefficient related to the taper properties\citep{Gorodetsky1999} and $K_e^0$ is the external coupling rate for the pendulum's equilibrium position, $d_0$. This can be expanded as a series of harmonics, $\cos(n\Omega_mt)$, where $n=1,2,3\ldots$. Since $\delta$ is generally small, $K_e(t)$ can be linearized to its first order term, corresponding to $\Omega_m$, i.e. $K_e(t)\approx K_e^0(1+2\beta\cos(\Omega_mt))$, where $\beta=I^0_1(\gamma\delta)$. $I^0_1$ represents the modified Bessel function of the first kind. Substituting $\Delta(t)$ and $K_e(t)$ into Eq. \ref{eq:Trans}, the Fourier transformed (FFT) transmission spectrum, $\tilde{T}$, at $\Omega_m$ is:
\begin{equation}
\begin{split}
\widetilde{T}\approx \frac{4K_e^0}{[(K_e^0+1)
\Delta_0]^2+\Delta_0^4+F(K_e^0)}\times \\
\left|2\Delta_0\Delta_g-2\beta({K_e^0}^2-\Delta_0^2-1)\right|.
\end{split}
\label{eq:FFTTrans}
\end{equation}
\begin{figure}
\includegraphics[width=9cm]{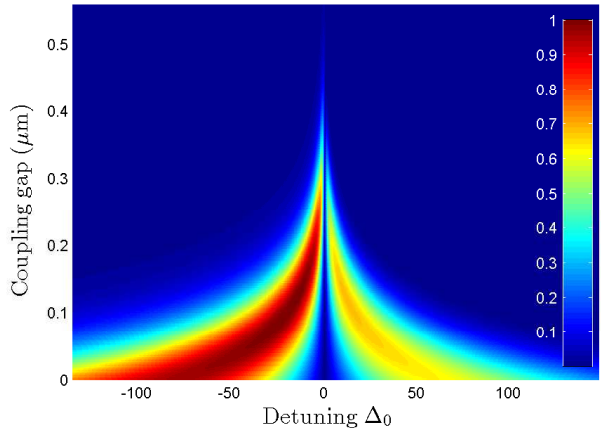}
\caption{\label{fig:MMSimulation} Numerically calculated transduction response of WGMs to the micropendulum motion. The color bar shows the 1 kHz mechanical mode relative peak amplitudes.}
\end{figure}
Here, $F(x)=x^4+4x^3+6x^2+4x+1$ is a polynomial expression. For conventional optomechanical systems\cite{Gavartin2012hybrid} the major contributor to the optomechanical transduction is dispersion and the effect of $\beta$ is normally omitted. Hence, the resulting spectrum, $\widetilde{T}$, is symmetric around  $\Delta_0=0$. In contrast, when $\Delta_g=0$, i.e. when the contribution from dispersion is negligible compared to that from dissipation, $\widetilde{T}$ produces a symmetric response around zero detuning again. 
In optomechanical systems such as ours, the motion-induced dissipation and dispersion  interfere. One can see from Eq. \ref{eq:FFTTrans} that the symmetry is broken if $\beta$ is non-negligible (typically like that shown in Fig.\ref{fig:ExpSetup}(e)).  Furthermore, since $K_e^0$ and $\Delta_g$ vary with $d_0$, the transduction spectrum changes with the coupling gap. Fig. \ref{fig:MMSimulation} is a numerical simulation of the transduction at different coupling gaps and detunings. This was obtained by directly solving Eq. \ref{eq:Trans} using the same parameters as in our experiments, which will be described later. It can be seen that, due to the two different mechanisms, the transduction signal should be higher when the laser is red-detuned relative to the WGM resonance. Also, if the micropendulum is closer to the fiber, the transducted signal  is stronger, with the strongest signal at critical coupling.
\begin{figure}
\centering
\includegraphics[width=5.6cm]{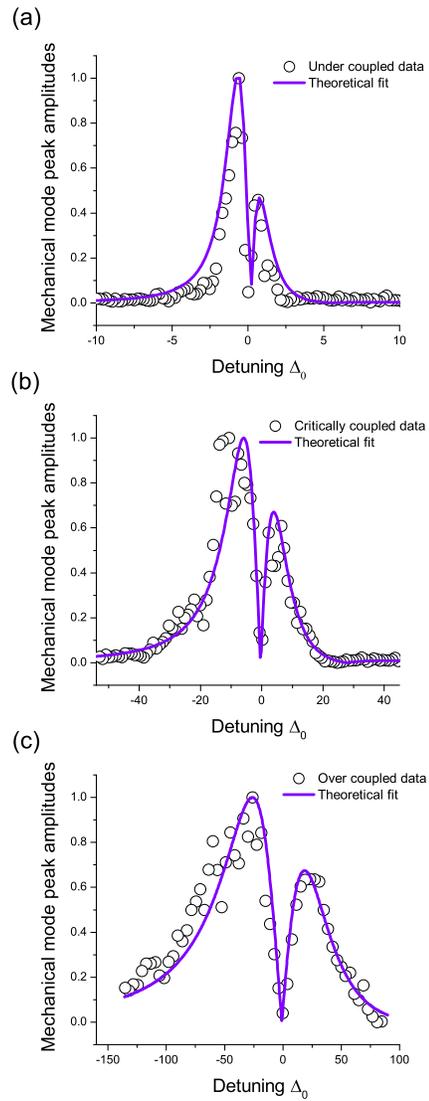}
\caption{Transduction response profiles. (a) undercoupled ($d_0=0.38$ $\mu$m), (b) critically coupled ($d_0=0.21$ $\mu$m), and (c) overcoupled ($d_0=0.12$ $\mu$m). Solid lines are theoretical fits. The experiments were repeated four times and data averages are presented.}
\label{fig:DetunedFit}
\end{figure}

Previously, in a similar experimental system, we observed that the signal peak was higher when the laser was red-detuned relative to the WGM resonance. This was attributed to thermal effects\cite{Wu2012}. For a high input power, a silica sphere can absorb light and generate heat, subsequently shifting the WGMs\cite{Carmon2004}. This is similar to the effect of dispersion on the system. In order to uncover the fundamental dissipative and dispersive mechanisms in the micropendulum system, we redid our earlier experiments\cite{Wu2012} using an input power of about 100 nW. The laser was scanned in steps of 0.6 MHz, spanning more than one full WGM linewidth, and FFT peaks were recorded for each value of $\Delta_0$. The measurement was repeated for different coupling gaps. The transduction response of the WGM to the micropendulum motion (i.e. the amplitude of the FFT peak) as a function of detuning was then reconstructed from the data. To isolate the system from environmental disturbances, the experiment was conducted in vacuum at 40 Pa. 
\begin{figure}[t]
\includegraphics[width=7.5cm]{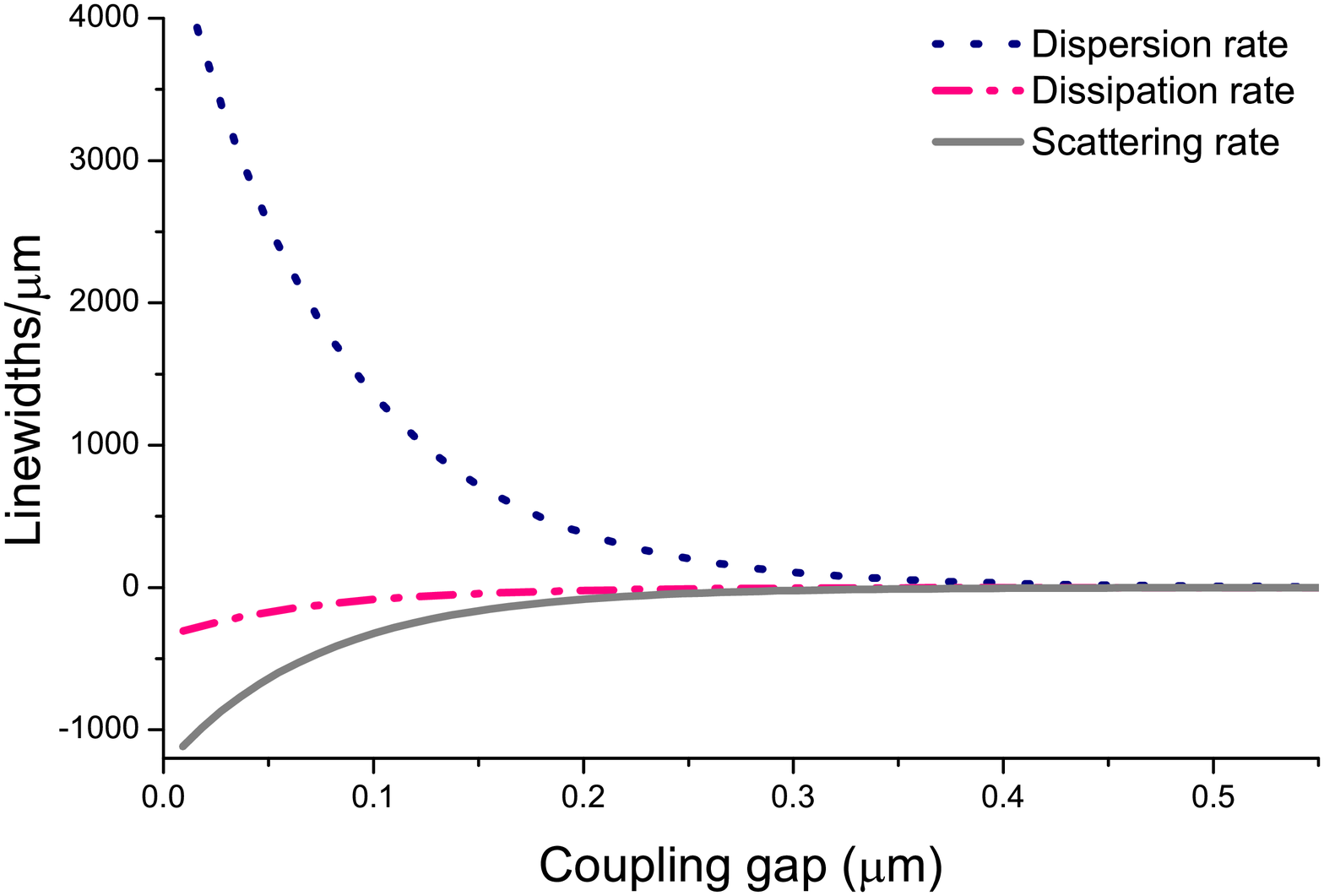}
\caption{\label{fig:RatesVSGap}The dispersion (dashed), dissipation (dot-dashed) and scattering (solid) rates as a function of coupling gap. The curves are fits taken from the experimental results.}
\end{figure}

\begin{figure}
\includegraphics[width=7.5cm]{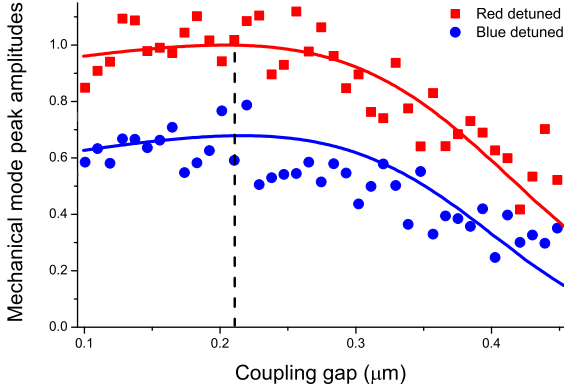}
\caption{\label{fig:MMDetunedGapFit} Transduction maxima for different coupling gaps. Red squares (blue dots) are experimental results for red- (blue-) detuned transduction maxima. The red (blue) solid line is the theoretical prediction. Critical coupling is represented as the dashed vertical line. The experimental data are the average of four experimental trials.}
\end{figure}

Typical response spectra are shown in Fig. \ref{fig:DetunedFit}. To further investigate the contributions of dispersion and dissipation, the experimental results were fitted with the discussed theory. The mechanical mode frequency was observed to be $\Omega_m/2\pi=1.05$ kHz and the accumulated oscillator displacement was estimated as $\approx 1$ nm at room temperature\cite{Wu2012}. By fitting the experimentally measured \textit{T}, at $\Delta_0=0$, for different coupling gaps, the equilibrium relation for $K_e^0$ was reconstructed as an exponentially decaying function of $d_0$, such that $K_e^0=A\exp[\gamma(d_0-B)]$ with fitting parameters $A=4.2$, $\gamma=-14$ $\mu\text{m}^{-1}$, and $B=0.13$ $\mu$m. Similarly, an independent experimental measurement was taken for the optical mode spectral position for different coupling gaps. The derivative yields a reconstructed equilibrium dispersion relation, which is an exponentially rising function of $d_0$, i.e. $\Delta_g=C\exp(\zeta d_0)$, where $C=4.5$ and $\zeta=-12$ $\mu$m$^{-1}$ are the fitting parameters. The intrinsic loss, $\kappa_0$, was estimated from the optical linewidth of the WGM when it was far undercoupled, so that $\kappa_0/2\pi\approx13$ MHz. $\kappa_s$ was taken into consideration in the theory model for a better fit. The dispersion, dissipation, and scattering rates, which are shown in Fig. \ref{fig:RatesVSGap}, increase with decreasing gap. Therefore, to increase the transduction signal, the micropendulum has to be positioned closer to the taper. In our case, maximum transduction can be reached at the critical coupling gap, as illustrated in Fig. \ref{fig:MMDetunedGapFit}. The ratio between the dispersion and external dissipation rates was also estimated and it was found to vary  weakly with the gap. Although the individual rates are relatively low, the ratio between the dispersion and dissipation rates is typically 20 fold compared to a 500 fold in the split-beam nanocavity\cite{Wu2014}. This indicates a significant improvement towards dissipation dominant optomechanics.

Here, the ratio between the scattering and external rates is about four, which means  scattering cannot be ignored. Since the ratio between the dispersion and external dissipation is almost constant for different gaps, one should expect a similar asymmetric lineshape in Fig. \ref{fig:DetunedFit}(a) and (b). However, experimental data show a weaker asymmetry. By considering the contribution of the scattering loss, such discrepancy can be explained. In a split-beam nanocavity\cite{Wu2014}, the ratio was shown to be 150 (including intrinsic and scattering rates). Compared to this, in our system, the external dissipation has a more significant contribution. Furthermore, the scattering loss can be eliminated by having an ideal taper\cite{Spillane2003}, so our system may have an advantage in the cooling scheme requiring external dissipative coupling\cite{Krause2012}. 

\begin{figure}[h]
\begin{center}
\includegraphics[width=8.6cm]{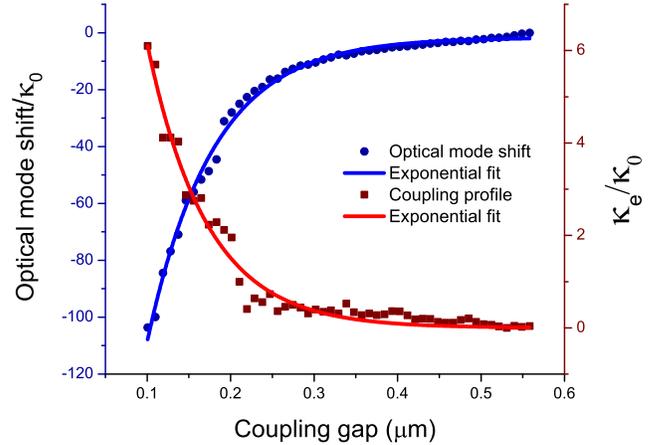}
\caption{The dispersion and dissipation profiles as a function of coupling gap.\label{fig:DetandKeVSGap}}
\end{center}
\end{figure}

The dispersion and dissipation profiles discussed previously were  obtained from the fitted experimental data. As shown in Fig. \ref{fig:DetandKeVSGap}, these are, $\Delta\omega/\kappa_0=-1.7-380\exp(-12d_0)$ and $K_e^0=25\exp(-14d_0)$.

The total decay rate, $K_t^0$ was deduced from the loaded $Q$-factor (given by $1/Q=1/Q_0+1/Q_e+1/Q_s$) measured for different coupling gaps. The measured optical line width is 13 MHz in far undercoupled regime. With $K_e^0$, the scattering rate can be calculated from $K_s^0=K_t^0-K_e^0-1$, which was fitted with an exponentially decaying profile as a function of $d_0$, similar to $K_e^0$ as $0.13+81\exp(-13.7d_0)$. $K_e(t)$, $\Delta(t)$ and $K_s(t)$ were updated with the above equilibrium functions. To account for the scattering profile in the model, they were then directly substituted into Eq. 2, which was used to plot Fig. 2. For Fig. 3, the coupling conditions varied from undercoupled (Fig.3(a)) to overcoupled (Fig.3(c)). We also plotted all maxima of the detuning response curves for coupling gaps ranging from 0.1 to 0.45 $\mu$m with the respective theoretical prediction curves presented in Fig. 5. The resolution of the nanopositioner is 10 nm, hence we allowed a gap tolerance of 20 nm in the theoretical fittings. All the data plots in Fig. 3 and 5 are averaged over four experimental trials for each gap.  Furthermore each data point for the critical (over) coupled plot (as in Fig. 3(b) and (c)) is an average of four (15) data points. 

The rates of the dispersion, dissipation, and scattering are derivatives of $d_0$\cite{Wu2014}, which are $4500 \exp(-12 d_0)$, $-346 \exp(-14 d_0)$, and $-1108 \exp(-13.7 d_0)$, respectively. These are plotted as a function of the gap in Fig. 4. Note that the \textit{y}-axis represents optical linewidth/$\mu$m. Table \ref{tab:ratesvsgap} gives the rates for cases shown in Fig. 3. As evident, the dispersion and dissipation effects on the system are similar for most of the coupling gaps, but  the overall transduction amplitude increases with decreasing gap. The ratio between the dispersion and dissipation rates goes typically as 20 fold compared to 500 fold in a split-beam nanocavity\cite{Wu2014}. Furthermore, these rates can be tuned by coupling to different positions of the fiber taper. In principle, nearly ideal coupling is achievable \cite{Spillane2003}, hence pure external dissipation dynamics can be realized.

\begin{table}[h]
\begin{center}
\includegraphics[width=8.6cm]{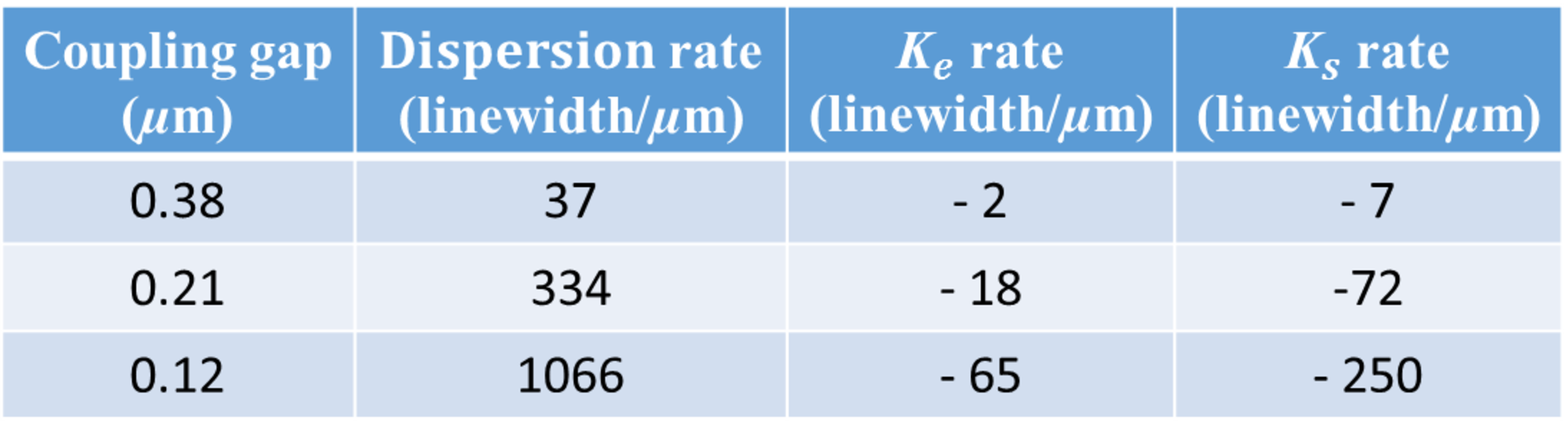}
\caption{The dispersion, dissipation and scattering rate parameters for under, critical, and over coupled regime of the system.\label{tab:ratesvsgap}}
\end{center}
\end{table}

In the fitting there are discrepancies, which are around $\pm 10$\% error. It may originate from the inaccuracy of the gap measurement. or excessive noise  in the low frequency FFT spectrum due to low (100 nW) detectable laser powers. However, it is still evident that an  asymmetric, coupling-gap-dependent shape in the transduction spectra exists in the practical micropendulum optomechanical system.

In conclusion, the relatively large mechanical oscillation of the micropendulum modulates the WGMs in both a dissipative and dispersive manner. We demonstrated, theoretically and experimentally, the interference of these mechanisms in a WGM optomechanical system. This system may be useful for optomechanical sensing\cite{Fan2015}. In future work, the dispersion may be further decreased in order to study dissipative cavity optomechanical cooling \cite{PhysRevLett.107.213604} and trapping \cite{Ward2009}.

This work was supported by funding from the Okinawa Institute of Science and Technology Graduate University. The authors thank Y. Zhang for discussions.

\bibliography{Pendulum}
\bibliographystyle{aipnum4-1}

\end{document}